\documentclass{article}
\usepackage{hiph-preprint}
\usepackage{epsfig}

\newcommand{\beq}{\begin{eqnarray}}
\newcommand{\eeq}{\end{eqnarray}}
\newcommand{\be}{\begin{eqnarray*}}
\newcommand{\ee}{\end{eqnarray*}}

\newcommand{\D}{{\cal D}}
\newcommand{\Pom}{{\hspace{ -0.1em}I\hspace{-0.25em}P}}
\def\lsim{\raise0.3ex\hbox{$<$\kern-0.75em\raise-1.1ex\hbox{$\sim$}}}
\def\gsim{\raise0.3ex\hbox{$>$\kern-0.75em\raise-1.1ex\hbox{$\sim$}}}
\volnumber{22} \issuenumber{1} \edyear{2005}                             
\frompage{000} \topage{000}                                              
\recrevdate{1 January 2005}                                              
\def\rightmark{Gluon shadowing and unitarity effects}
\def\leftmark{}
 
\title{Gluon shadowing and unitarity effects} 
\authors{ 
{K.~Tywoniuk$^1$, I.~C.~Arsene$^1$, L.~Bravina$^1$, A.~B.~Kaidalov$^2$ and E.~Zabrodin$^1$ %
\index{Tywoniuk, K.} 
\index{Arsene, I. C.} 
\index{Bravina, L.}
\index{Kaidalov, A. B.}
\index{Zabrodin, E.}
}\\[2.812mm]
{\normalsize
\hspace*{-8pt}$^1$ Department of Physics, University of Oslo\\ 
0316 Oslo, Norway\\[0.2ex] 
\hspace*{-8pt}$^2$ Institute of Theoretical and Experimental Physics\\ 
117259 Moscow, Russia
}}
 
\abstract{New data from HERA experiment on deep inelastic scattering have been
used to parametrize nucleon and Pomeron structure functions. Within
the Gribov theory, the parameterizations were employed to
calculate gluon shadowing for
various heavy ions. The latter was compared with predictions from
other models. Calculations of multiplicity reduction due to gluon
shadowing for d+Au
collisions at forward rapidities at $\sqrt{s_{NN}}$=200 GeV are in
good agreement with BRAHMS data on the nuclear modification factor.}

 
\makeindex
\begin{document}
 
\maketitle

\section{Introduction}\label{intro}
For low energies of the incoming beam in a hadron-nucleus collision,
the successive elastic rescatterings of the initial hadron on the
various nuclei of the nucleus are well described within the
probabilistic Glauber model \cite{glauber}. At higher energies,
corresponding to $E_{crit} \sim m_{\scriptstyle{N}} \mu R_A$, the
hadronic fluctuation length can become of the order of nuclear
radius and there will be coherent interaction of constituents of the
hadron with several nucleons of the nucleus. Within the Gribov
approach \cite{gribov1}, this corresponds to summing up
contributions of inelastic intermediate states, and leads to a
reduction of the total cross section of the reaction, i.~e. to nuclear
shadowing.

We calculate the total amount of gluon shadowing for low values
of the Bjorken variable $x$ for heavy ions, ignoring for
the time being the contribution from the quarks. The most recent data on
diffractive structure functions are used and much stronger shadowing
effects than previously expected are found. These effects will lead to
a strong multiplicity reduction in A+A collisions at RHIC
and LHC energies.

\section{The Model}
The diffractive $\gamma^* N$ scatterings are described by Pomeron
exchange. The scattering amplitude of an incoming photon with
virtuality $Q^2$ on a nuclear target, consisting of A nucleons, can
then be written as \cite{cap98}
\beq
\label{eq:sum}
\sigma_A \;=\; A\sigma_N \,+\, \sigma_A^{(2)} \,+\, ...\;.
\eeq
The second term in (\ref{eq:sum}) is negative and is related to
diffractive DIS through the AGK cutting rules \cite{agk}. Higher order
rescatterings in (\ref{eq:sum}) are model dependent. The Schwimmer
unitarization \cite{schwimmer} for the $\gamma^* A$ cross
section is used to obtain the shadowing ratio
\beq
\label{eq:sch}
R^{Sch}_{\scriptscriptstyle{A/N}} (x) \,\equiv \,
    \frac{\sigma_{\gamma* A}^{Sch}}{A \,\sigma_{\gamma^* 
    \scriptscriptstyle{N}}}  \;=\; \int 
\mbox{d}^2b \; \frac{T_A (b)}{1 \,+\, (A-1) f(x, Q^2) T_A (b) } \;,
\eeq
where $f(x,Q^2)$ is the effective shadowing function, $T_A(b)$ is the
nuclear density profile normalized to unity and standard DIS
variables are used. Following \cite{cap01,H1abs} in choice of
parameters and  kinematics, one can get the shadowing function as 
\beq
\label{eq:fsimpl}
f(x, Q^2) \;&=&\; 4\pi\; \int_x^{x_\Pom^{max}} \mbox{d}x_\Pom \,B(x_\Pom)\,
\frac{F_{2 \D}^{(3)}(x_\Pom, Q^2, \beta)}{F_2 (x, Q^2)} \; F_A^2
(t_{min.}) \;.
\eeq
Here $B(x_\Pom) = 0.184 - 0.02 \ln \left(x_\Pom \right)\,\mbox{fm}^2$, and $F_A$ is
the form factor of the nucleus.

Calculations are made both for $x_\Pom^{max}
= 0.1$ as in \cite{cap98} and for $x_\Pom^{max} = 0.03$ as in
\cite{fgs03}.
The structure functions $F_2 (x,Q^2)$ and 
$F_{2 \D}^{(3)} (x_\Pom, Q^2, \beta)$ are determined from
experiment. At small $x$, gluon shadowing is found to be dominant.
Quark contribution to the structure functions is not considered in
what follows.
Shadowing due to quarks, obtained wihtin the same approach, was discussed in
\cite{cap98}.

The gluon parton distribution functions (PDF) for nucleon and Pomeron were
measured at intermediate $Q^2$ at the HERA experiments ZEUS and H1,
correspondingly. The next to leading order (NLO)
ZEUS-S QCD fit for the gluon PDF of the nucleon \cite{chekanov03} at
$Q^2 = 7 \mbox{ GeV}^2$, and
the gluon PDF for the Pomeron (diffractive structure function)
\cite{H1abs} at $Q^2 = 6.5 \mbox{ GeV}^2$ were both parametrized by
\beq
x \, g(x, Q^2) \;=\;  A x^{-\delta} \left(1-x \right)^\gamma \;,
\label{eq:fitfunc}
\eeq
where the fitting parameters $\left\{A,\delta,\gamma \right\}= \left\{
1.9,0.19,6.7 \right\}$ were obtained for the nucleon and
$\{0.38,0.28,0.17 \}$ for the Pomeron case, respectively. The $Q^2$-dependence of the fitting parameters is weak for moderate $Q^2$, and so we neglect it for the sake of simplicity.

\section{Numerical results}\label{result}
\begin{figure}[t]
  \begin{minipage}[t]{0.5\linewidth}
    \centering
    \includegraphics[width=\linewidth]{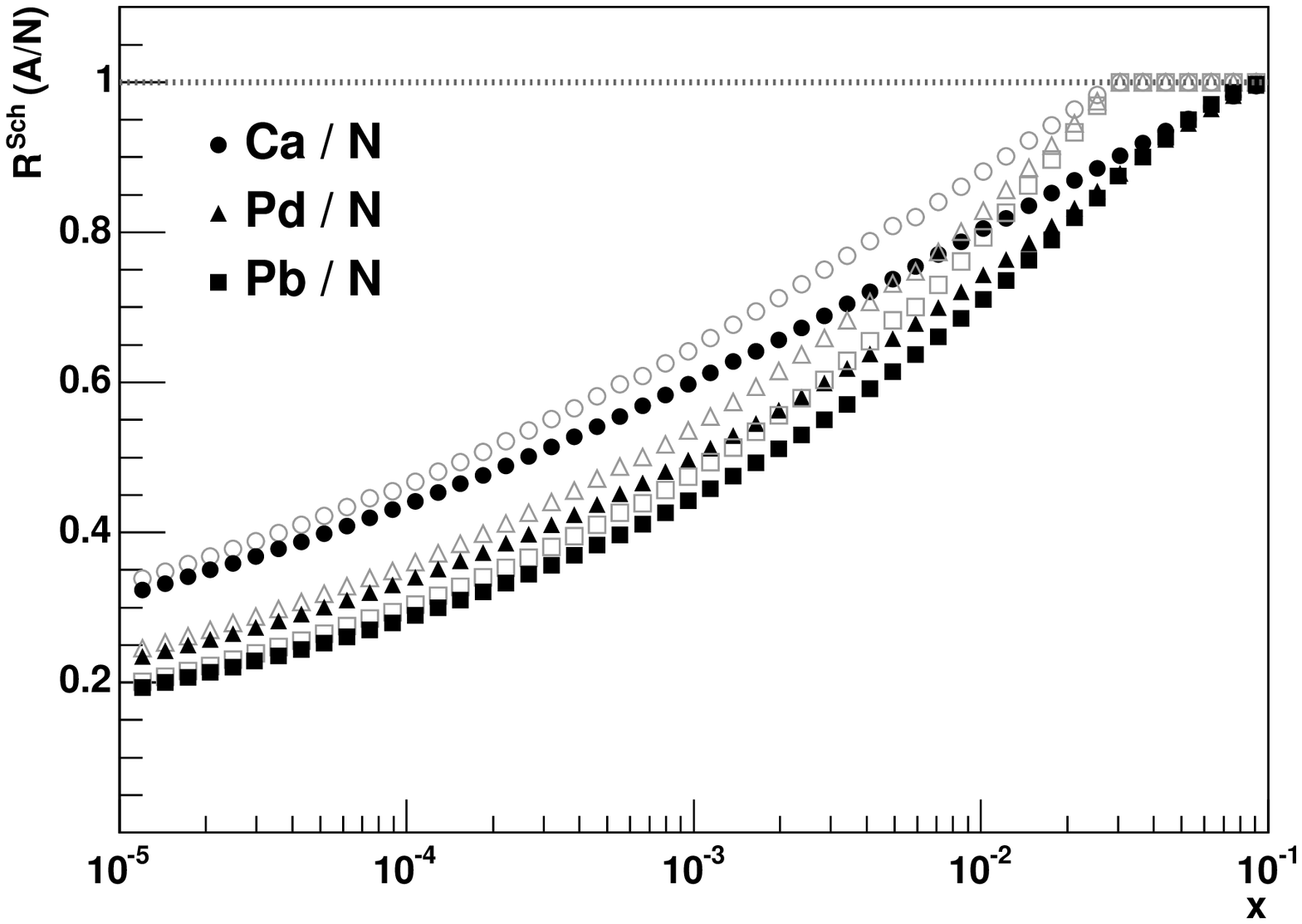}
    \caption{Gluon shadowing for heavy ions. Closed (open) symbols are for
      $x_\Pom^{max} = 0.1$ ($0.03$).}
    \label{fig:AN}
  \end{minipage}
  \begin{minipage}[t]{0.5\linewidth}
    \centering
    \includegraphics[width=\linewidth,height=1.81in]{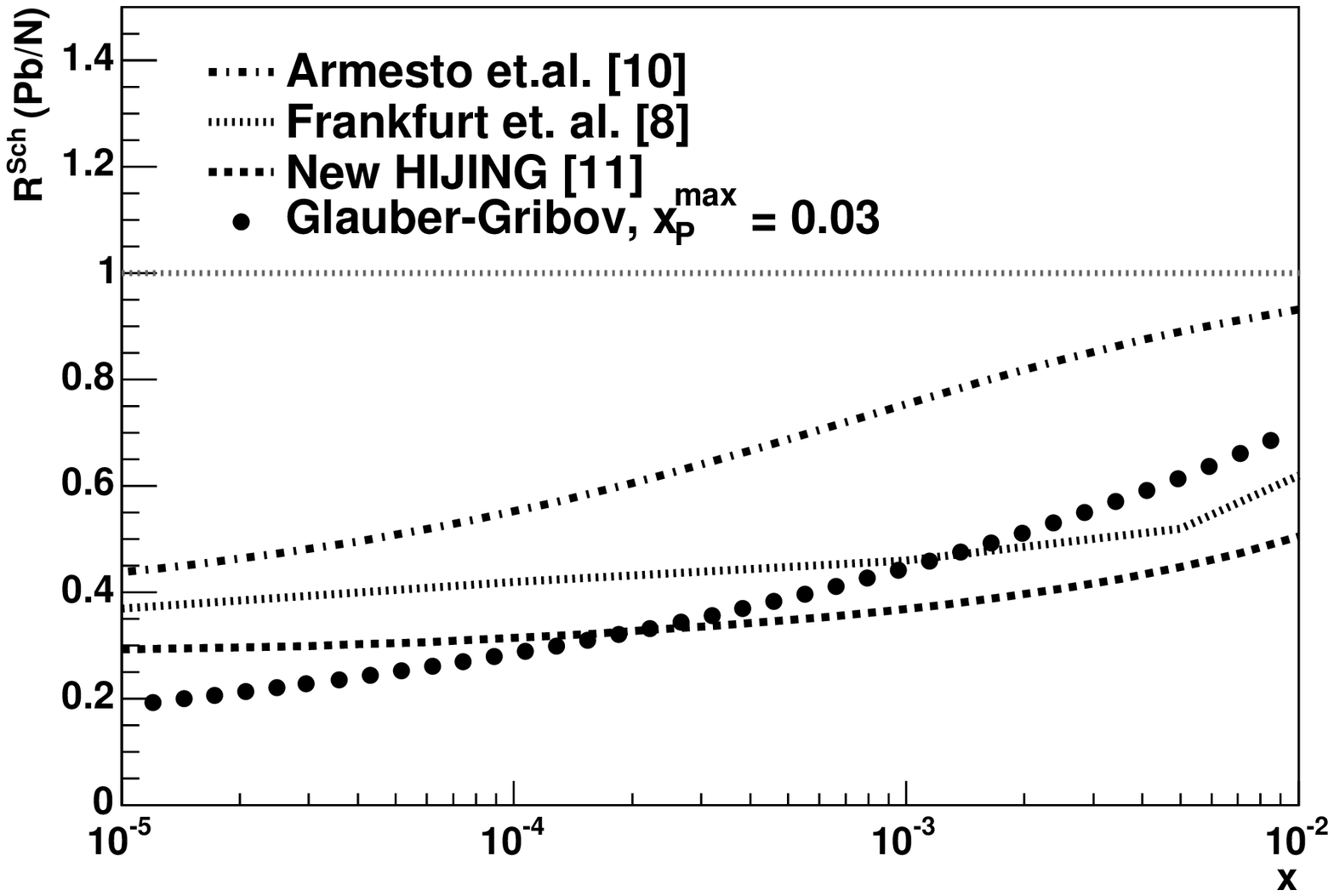}
    \caption{Comparison of theoretical predictions for the Pb/nucleon
    ratio, at fixed $Q^2$.}
    \label{fig:comp}
  \end{minipage}
\end{figure}

Gluon shadowing for various heavy ions (Ca, Pd and Pb) from (\ref{eq:sch})
is presented in Fig.~\ref{fig:AN}. The gluon shadowing is very strong
at small $x$, and disappearing at $x = x_\Pom^{max}$. This is a
consequence of the coherence effect in the form factor,
and the vanishing integration domain in (\ref{eq:fsimpl}). Gluon
shadowing is as low as 0.2 for the Pb/nucleon ratio. 

A comparison of our results for Pb/nucleon ratio at 
$Q^2 = 6.5 \mbox{ GeV}^2$ with $x_\Pom^{max} = 0.03$, with those of
others, calculated at $Q^2 = 
5 \mbox{ GeV}^2$ is presented in Fig.~\ref{fig:comp}. For $x \leq
10^{-3}$ our model predicts the stronger gluon shadowing compared to
\cite{armesto02} (dashed-dotted line) and \cite{fgs03}
(dotted line), while \cite{hijing} (dashed line)
predicts the strongest effect down to $x \sim 10^{-4}$.

\section{Shadowing effects in d+Au collisions}
The model is now employed to study multiplicity reduction in d+Au collisions
at ultra-relativistic energies. Deuteron is treated as a point-like
particle in impact parameter space, but with the shadowing
found from (\ref{eq:sch}). The collision is described by two-jet
kinematics through $ x_{p (t)} \;=\; c\,
p_T \, e^{\pm \eta}/ \sqrt{s} $, where $p_T$ is the transverse
momentum of the particle, and fixed at $Q^2$. We assume that most of the
high-$p_T$ particles come from jets $c$ times more energetic than the
measured one. The theoretical prediction is given by \cite{cap99}
\beq
\label{eq:shadow}
R^{theo}_{d+Au} \;&=&\; R^{Sch}_d (x_{p}) R^{Sch}_{Au} (x_{t}) \;.
\eeq
From here one obtains the multiplicity reduction due to shadowing
compared to the Glauber model. Then the model predictions for nuclear
modification factor (NMF) at forward rapidities are compared to BRAHMS
data at $\sqrt{s_{\scriptstyle{NN}}} = 200$ GeV \cite{brahms}. 

\begin{figure}[t]
  \begin{minipage}[t]{0.5\linewidth}
    \centering
    \includegraphics[width=1.1\linewidth]{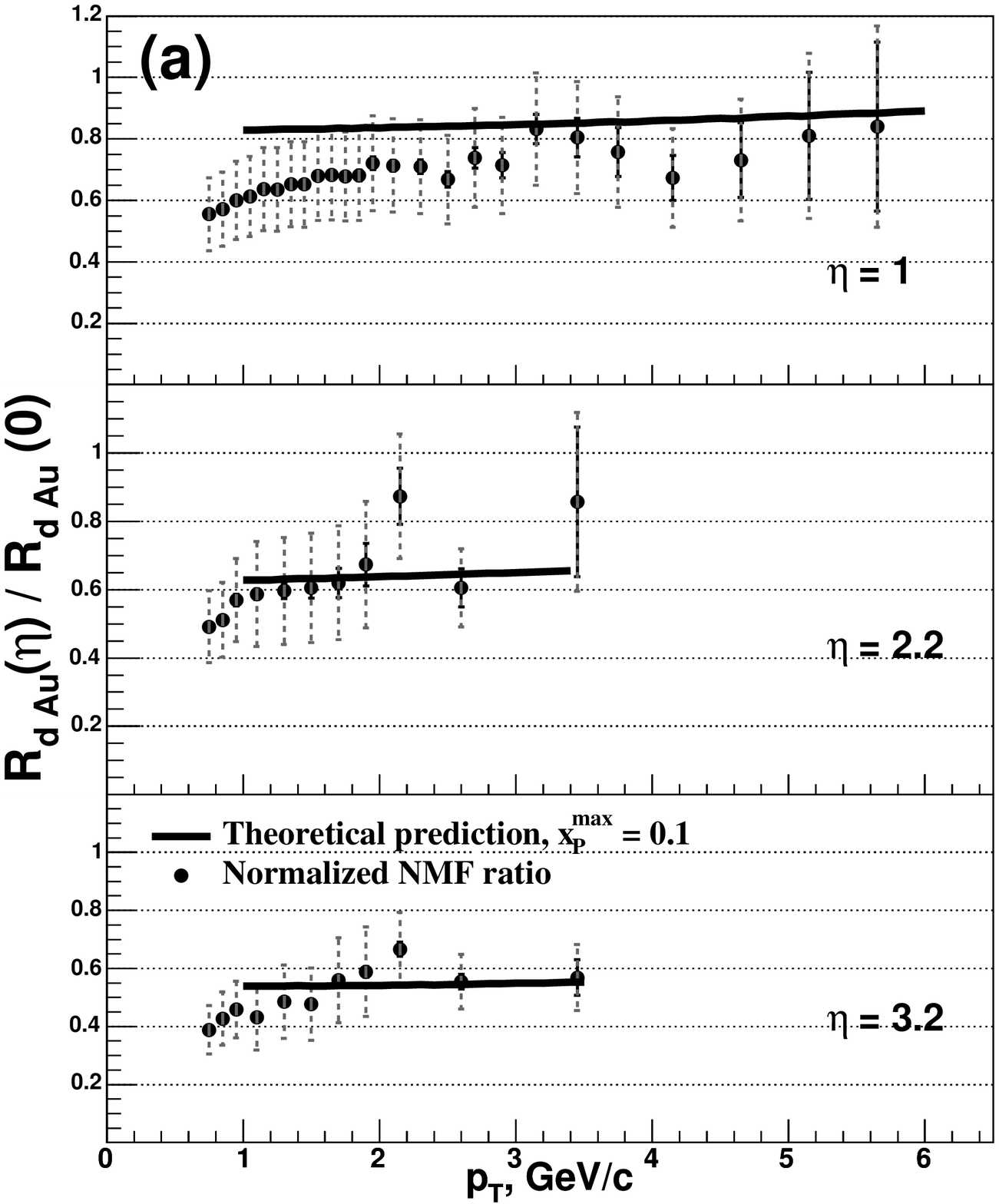}
  \end{minipage}
  \begin{minipage}[t]{0.5\linewidth}
    \centering
    \includegraphics[width=1.1\linewidth]{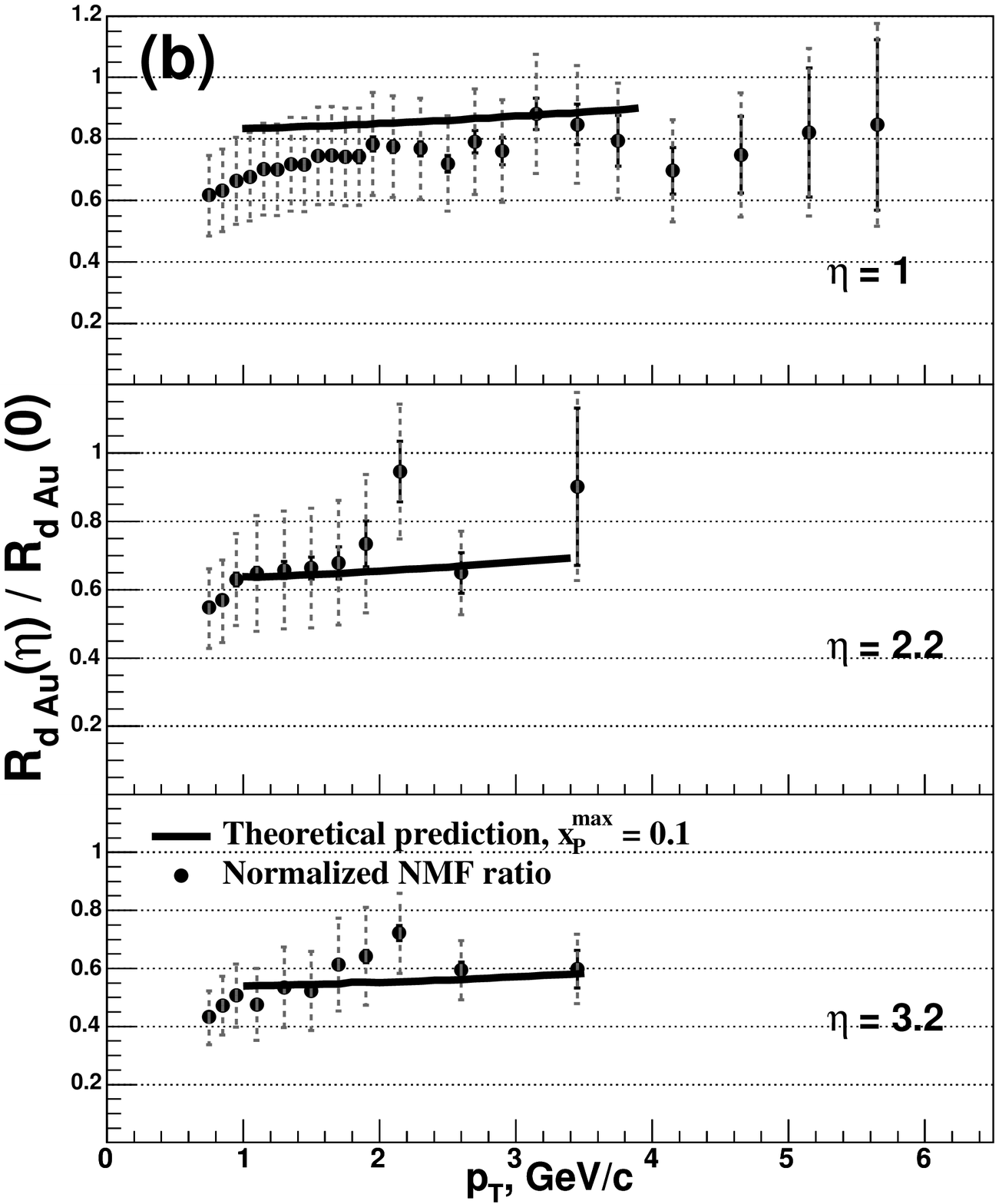}
  \end{minipage}
  \caption{NMF ratio for (a) $c = 3$ and (b) $c = 5$. See text for
  details.}
  \label{fig:res_c}
\end{figure}

We exclude the gluon shadowing effects in the BRAHMS NMF at $\eta =
0$ by defining $R^{\mathit{norm}}_{d Au} =
\left[R_{dAu}^{\mathit{exp}} / R_{d-Au}^{\mathit{theo}} \right]_{\eta
  = 0}$. The multiplicity reduction due to shadowing effects will
appear when we compare the NMF at forward rapidities, $\eta = 1,
\,2.2, \,3.2$ to $R^{\mathit{norm}}_{d Au}$. The ratio $\tilde{R}
\;=\; \left[R^{exp}_{d Au}\right]_\eta / R^{\mathit{norm}}_{d Au}$ is
plotted in Fig. \ref{fig:res_c} together with the predictions of
(\ref{eq:shadow}) for two different values of the parameter $c$. 
Statistical errors are denoted by the thick solid line, while the
systematic and statistical errors added up are denoted
by the dashed line. Cronin effect is assumed to be rapidity independent and is
cancelled out in the ratio.
The choice of $c$ does not affect the
result. Within the presented framework, one can conclude that suppression of the nuclear modification
factor at forward rapidities is mostly due to gluon shadowing in the
nuclei.
\linebreak
\linebreak
{\it{Acknowledgements:}} The authors acknowledge support from NFR and
\linebreak QUOTA program.


\vfill\eject
\end{document}